\documentclass[12pt]{article}               
\usepackage{a4}
\usepackage{epsfig}
\usepackage{amsmath,amssymb}

\def\un{{\underline 1}}
\def\et{{\underline \epsilon}}

\def\ezero{\mbox{${\bf e_0}$}}

\def\khat{\mbox{${\rm\bf\hat k}$}}

\def\phat{\mbox{${\rm\bf\hat p}$}}
\def\shat{\mbox{${\rm\bf\hat s}$}}
 
\def\er{\mbox{ ${\rm\bf r}$}}

\def\chat{\mbox{${\rm\bf\hat c}$}}
\def\xhat{\mbox{${\rm\bf\hat x}$}}
\def\yhat{\mbox{${\rm\bf\hat y}$}}
\def\zhat{\mbox{${\rm\bf\hat z}$}}

\begin{document}
\vspace*{-1.8cm}
\begin{flushright}
{\bf LAL 03-55}\\
\vspace*{0.1cm}
{ October 2003}\\
\end{flushright}

\vspace*{0.3cm}
\begin{center}
{\Large \bf Surfaces roughness effects on the transmission of Gaussian beams
 by anisotropic parallel plates}\\

\vspace*{0.5cm}
{\bf  F. Zomer}\\
\vspace*{0.25cm}
{\small \bf Laboratoire de l'Acc\'el\'erateur Lin\'eaire, IN2P3-CNRS}\\
{\small et Universit\'e de Paris-Sud, F-91405 Orsay cedex, France.} 
\end{center}
\vspace*{0.3cm}

\begin{abstract}
Influence of the plate surfaces roughness in precise ellipsometry experiments
 is studied. The realistic case of a Gaussian laser beam crossing a uniaxial platelet 
 is considered.
 Expression for the transmittance is determined using the first order perturbation theory.  
 In this frame, it is shown that interference takes place between the specular transmitted beam
 and the scattered field. This effect is due to the angular distribution of the Gaussian beam
 and is of first order in the roughness over wavelength ratio.
 As an application, a numerical simulation of the effects of quartz roughness surfaces 
 at normal incidence is provided. 
 The interference term is found to be strongly connected to the random nature of the surface
 roughness.
\end{abstract}


\section{Introduction}

 The high-accuracy universal polarimeter (HAUP) \cite{kobayashi} has proved to
 be a very useful instrument to measure the crystal optical properties
 (see for instance \cite{moxon,ortega,hernandez} and references therein).
 The principle is simple and was introduced a long time ago 
 (see \cite{bruhat} for an historical introduction):
 the light intensity measured after a rotating high quality
 polariser, a crystal plate (the sample)
 and a high quality rotating analyser, is fitted to a theoretical formula with
 several coefficients as free parameters where the delay due to birefringence 
 and optical activity can be determined.

 The accuracy of this instrument has now reached
 the few $10^{-5}$ level and systematic errors 
 contributing at this order of magnitude 
 have been investigated \cite{kremers,simon,folcia}. The conclusion is that
 roughness is most likely one of the main source of systematic 
 uncertainties.  
 
However, despite an extensive literature on surface
 roughness \cite{livre1,livre2}, no theoretical 
 expression for the transmission of a Gaussian beam by an anisotropic rough
 platelet is available. It is the purpose of this article to provide this
 expression. We consistently take into account the Gaussian nature of the laser
 beam, the multiple reflection inside the plate and the roughness of both faces  of the plate. To simplify the calculations we further restrict ourselves to 
 uniaxial homogeneous crystals. As a result, we find that
 unlike plane waves, specular Gaussian beams are affected by the 
 surfaces roughness, even in the first order perturbation
 theory.

 The physical origin of this phenomenon is the angular distribution, or plane wave expansion, 
 of Gaussian beams \cite{siegman}. Plane waves constituting a Gaussian beam having
 different wave vectors, a given plane wave can then be scattered in the specular direction  
 of the other ones. The resulting interference pattern
 leads to an a priori non vanishing contribution of the scattered field
 in the specular region. To some extent, this phenomenon is thus related to the
 near-specular scattering by rough surfaces introduced in \cite{nee}.
 
Another aspect of realistic platelet surfaces is the interface parallelism default.
 Depending on the wedge angle, this default can compete with roughness in the modifications
 of the transmitted beam polarisation. The nature of these effects is however different. 
 Given the relative orientation of the two plate interfaces, the wedge
 effect is univocal whereas roughness, as it will be shown in this paper, is of random nature. 
 It is then most likely that these two effects cannot compensate each other.
 In principle, the perturbative calculations
 reported in the present article holds for both effects. Nevertheless, the boundary matching method, 
 applied a long time ago to isotropic wedges \cite{brossel}, can be used to describe the wedge
 effect. We shall report this calulation in a future publication and restrict ourselves
 here on platelet roughness. 

 This article is organised as follows.
 In section \ref{theory} we derive the theoretical expressions and
 numerical simulations of quartz plates are presented in section \ref{numerical}. 

\section{Formalism}\label{theory}

The choice of the theoretical formalism is driven by the properties
 of the crystal plate surfaces under study. Fortunately, an exhaustive
 experimental study on crystal surfaces has recently
 been published \cite{surf}. Most of the high quality polished
crystal surfaces used in optics have a profile surface
 correlation length of the
 order of the optical wavelength and a root mean square roughness
 of the order of a few angstr\oe m. It means that  
 one can safely use a first order perturbation theory  \cite{acoustic} neglecting the
 local field effects\cite{local}. The more suitable
 formalism for our problem is the one introduced in \cite{hill} and 
 generalised to anisotropic overlayers in \cite{celli}. However, in 
 the latter reference, the anisotropy is treated perturbatively and only the
 reflection of plane waves is considered.
We shall then extend this formalism to platelet's transmission taking fully
 into account the plate anisotropy and treating
 perturbatively the plate roughness.

\begin{figure}[htbp]
\begin{center}
\epsfig{file=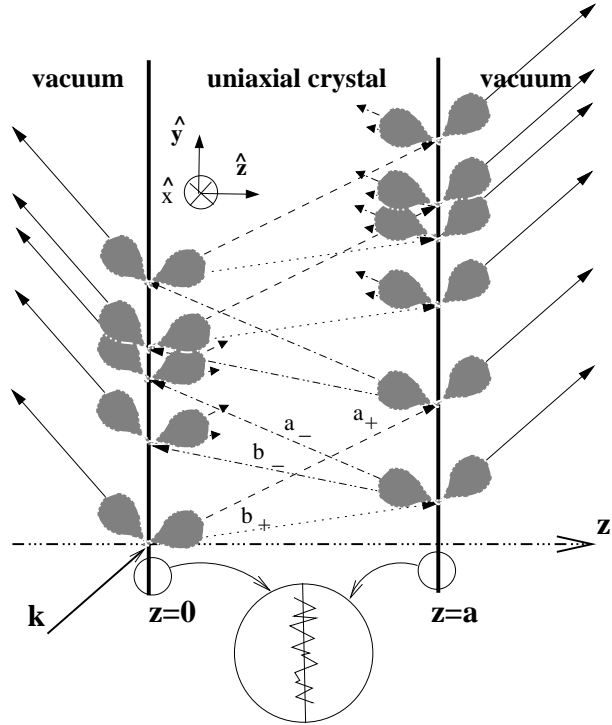,width=8cm}
\end{center}
\caption{Schematic view of the plane wave propagation in the anisotropic
 slab. For the sake of clarity, some of the inner reflected rays 
are represented by small arrows. The plane of
 incidence coincides with the plane $yz$. Symbols $a_\pm$ and
 $b_\pm$ correspond to the four possible propagation directions inside
 the medium. The vector basis  $\{\xhat,\yhat,\zhat\}$
 used throughout this article is also shown. The grey areas 
 symbolise the scattered light due to surfaces roughness.
    \label{fig1}}
\end{figure}

 In the following we tried to be concise, referring to
 \cite{hill,celli} for further details.   
 The wave equation corresponding to the system represented in figure \ref{fig1}
 is:
\begin{equation}\label{wave-eq}
\bf{\nabla}\times\bf{\nabla}\times \bf{E}(\bf{r})
=\omega^2\mu_0\bf{D}(\bf{r})
\end{equation}
with $\bf{D}(\bf{r})={\cal E}(z)\bf{E}(\bf{r})$ and
\begin{eqnarray}\label{epsil}
{\cal E}(z)=\Theta\biggl(-z+h_0(x,y)\biggr)\epsilon_0\un+
\biggl[\Theta\biggl(z-h_0(x,y)\biggr)
-\Theta\biggl(z-a-h_a(x,y)\biggr)\biggr]\et\nonumber\\
+\Theta\biggl(z-a-h_a(x,y)\biggr)\epsilon_0\un
\end{eqnarray}
where $\un$ is the $3\times 3$ identity matrix and $\Theta$ is the Heaviside 
 function. For uniaxial media, it is useful to write \cite{uniaxe}
 \[
\et=\epsilon_{\perp}
\un+(\epsilon_{\parallel}-\epsilon_{\perp})\chat\chat
\] with $\chat^T=\{c_x,c_y,c_z\}$ 
 the unit vector along the optical axis, $\chat\chat$ a Dyad and 
 $\epsilon_\perp=n_o^2\epsilon_0$, $\epsilon_\parallel=n_e^2\epsilon_0$
  the ordinary and extraordinary components of the dielectric tensor. 
 In equation (\ref{epsil}), the two functions $h_0(x,y)$ and $h_a(x,y)$ are the
 profiles of the two surfaces located at $z=0$ and $z=a$ respectively.
 As usual \cite{bennett}, we assume that the two planes $z=0$ and $z=a$ are 
 defined such that the mean profiles vanish, i.e. $<h_0>=<h_a>=0$.


  The solution of equation (\ref{wave-eq}) can be written $\bf{E}(\bf{r})=
\bf{E}_0(\bf{r})+\bf{E'}(\bf{r})$ with $\bf{E}_0(\bf{r})$ given
by the zero order wave equation
\begin{equation}\label{wave-eq0}
\bf{\nabla}\times\bf{\nabla}\times \bf{E}_0(\bf{r})
=\omega^2\mu_0\bf{D}_0(\bf{r})
\end{equation} 
where ${\bf D_0}({\bf r})={\cal E}_0(z){\bf E_0}({\bf r})$ and
\begin{equation}\label{epsil0}
{\cal E}_0(z)=\Theta(-z)\epsilon_0\un+
\biggl(\Theta(z)-\Theta(z-a)\biggr)\et
+\Theta(z-a)\epsilon_0\un.
\end{equation}
To first order in $\omega h/c$ \cite{celli},
 one has ${\cal E}(z)={\cal E}_0(z)+
\delta{\cal E}(z)$ with
\begin{equation}\label{dirac}
\delta{\cal E}(z)\approx
\biggl(h_a(x,y)\delta(z-a)-h_o(x,y)\delta(z)\biggr)
\biggl[
(\epsilon_\perp-\epsilon_0)\un+(\epsilon_{\parallel}-\epsilon_{\perp})
\chat\chat\biggr]
\end{equation}
with $\delta(z)$ the Dirac distribution.

To derive the differential equation for the first order scattered field
 $\bf{E'}(\bf{r})$, we introduce the Fourier transform
\[
{\bf E}({\bf K};z)={\cal F}[{\bf E}({\bf r})]=
\frac{1}{2\pi}\int{\bf E}({\bf r})\exp(i{\bf K}\cdot{\bf R})
{\rm d}^2{\bf R}
\] 
where ${\bf R}=({\bf r}\cdot\xhat)\xhat+({\bf r}\cdot\yhat)\yhat$ and 
${\bf K}=({\bf k}\cdot\xhat)\xhat+({\bf k}\cdot\yhat)\yhat$ with ${\bf k}$
 the wave vector. Here, since we are considering Gaussian beams, no spatial
 length is introduced in the Fourier transformation.

Taking the Fourier transform of equations (\ref{wave-eq}) and (\ref{wave-eq0}) and then
 subtracting them, one obtains\cite{celli}
\begin{equation}\label{gen-wave-eq}
\biggl(-i{\bf K}+\zhat\frac{\partial}{\partial z}\biggr)
\biggl(-i{\bf K}\cdot{\bf E'}({\bf K};z)+\frac{\partial E'_z({\bf K};z)}
{\partial z}\biggr)
-\biggl(-K^2+\frac{\partial^2}{\partial z^2}\biggr){\bf E'}({\bf K};z)
=\omega^2\mu_0{\bf D'}({\bf K};z),
\end{equation}
with ${\bf D'}({\bf K};z)={\bf D}({\bf K};z)-{\bf D}_0({\bf K};z)$. For
 perturbative stability, the wave equation must be written as a function
 of the continuous electric field components \cite{hill},
 that is $E'_x$, $E'_y$ and $D'_z$. We shall do it separately for the 
$s$ and $p$ scattered waves as in \cite{hill}. However,
 before providing the solutions we introduce\cite{celli} the following
 useful vector function: 
\begin{equation}\label{fdef}
{\bf F}({\bf r})={\bf D}({\bf r})-{\cal E}_0(z){\bf E}({\bf r})
=\delta{\cal E}(z){\bf E}({\bf r})
\Leftrightarrow {\bf F}({\bf r})=
{\bf D'}({\bf r})-{\cal E}_0(z){\bf E'}({\bf r}),
\end{equation}
which gathers the infinitesimal contributions to the perturbated wave equation.
 To first order, one gets:
\begin{eqnarray}
F_x({\bf r})\approx\frac{1}{{\cal E}_{zz}(z)}\biggl(&
\delta{\cal E}_{xz}(z)D_z(\er)+\biggl[{\cal E}_{zz}(z)\delta{\cal E}_{xx}(z)
-{\cal E}_{xz}(z)\delta{\cal E}_{xz}(z)\biggr]E_x(\er)+\nonumber\\
 &\biggl[{\cal E}_{zz}(z)\delta{\cal E}_{xy}(z)
-{\cal E}_{yz}(z)\delta{\cal E}_{xz}(z)\biggr]E_y(\er)
\biggr)\,,\label{Fvec}\\
F_y({\bf r})\approx\frac{1}{{\cal E}_{zz}(z)}\biggl(&
\delta{\cal E}_{yz}(z)D_z(\er)+\biggl[{\cal E}_{zz}(z)\delta{\cal E}_{xy}(z)
-{\cal E}_{xz}(z)\delta{\cal E}_{yz}(z)\biggr]E_x(\er)+\nonumber\\
 &\biggl[{\cal E}_{zz}(z)\delta{\cal E}_{yy}(z)
-{\cal E}_{yz}(z)\delta{\cal E}_{yz}(z)\biggr]E_y(\er)
\biggr)\,,\\
F_z({\bf r})\approx\frac{1}{{\cal E}_{zz}(z)}\biggl(&
\delta{\cal E}_{zz}(z)D_z(\er)+\biggl[{\cal E}_{zz}(z)\delta{\cal E}_{xz}(z)
-{\cal E}_{xz}(z)\delta{\cal E}_{zz}(z)\biggr]E_x(\er)+\nonumber\\
 &\biggl[{\cal E}_{zz}(z)\delta{\cal E}_{yz}(z)
-{\cal E}_{yz}(z)\delta{\cal E}_{zz}(z)\biggr]E_y(\er)
\biggr)\,.\label{Fvec-fin}
\end{eqnarray}
where ${\cal E}_{ij}(z)$ and and $\delta{\cal E}_{ij}(z)$
 are the components of the symmetric dielectric
 tensors of equations (\ref{epsil}) and (\ref{dirac}). In the leading order
 perturbation theory, one further set\cite{hill}
 $E_x(\er)\approx E_{0x}(\er)$,
 $E_y(\er)\approx E_{0y}(\er)$ and
 $D_z(\er)\approx D_{0z}(\er)$
 in equations (\ref{Fvec}-\ref{Fvec-fin}).
\subsection{$p$ scattered wave}
Projecting equation (\ref{gen-wave-eq})
 onto $\zhat$ and utilising ${\bf \nabla}\cdot{\bf D}=0$ and equation (\ref{fdef}),
 we obtain the $p$  wave equation: 
\begin{align}
 &-\frac{\partial}{\partial z}\frac{1}{\epsilon_0(z)}
\frac{\partial D'_z({\bf K};z)}{\partial z}-\omega^2\mu_0 D'_z({\bf K};z)
+\nonumber\\
 &i({\bf K}\cdot\chat)\frac{\partial}{\partial z}
 \biggl[\frac{\Delta(z)}{\epsilon_0(z){\cal E}_{zz}(z)}
 \biggl(c_z D'_z({\bf K};z)+\epsilon_0(z)\biggl[c_x E'_x({\bf K};z)+
 c_y E'_y({\bf K};z)\biggr]\biggr)\biggr]+\nonumber\\
 &\frac{K^2}{{\cal E}_{zz}(z)}\biggr(D'_z({\bf K};z)
-{\cal E}_{xz}(z) E'_x({\bf K};z)-{\cal E}_{yz}(z) E'_y({\bf K};z)
\biggr)
\nonumber\\
&=\frac{\partial}{\partial z}\frac{1}{\epsilon_0(z)}
\biggl(-i{\bf K}\cdot{\bf F}({\bf K};z)+
i\frac{\Delta(z)}{{\cal E}_{zz}(z)}({\bf K}\cdot\chat)c_zF_z({\bf K};z)
\biggr)\nonumber\\
&+\frac{K^2}{{\cal E}_{zz}(z)}F_z({\bf K};z)
\label{pert-p}
\end{align}
where we introduced
 $\epsilon_0(z)=\biggl[\Theta(-z)+\Theta(z-a)\biggr]\epsilon_0
+\biggl[\Theta(z)-\Theta(z-a)\biggr]
\epsilon_{\perp}$ and
 $\Delta(z)=\biggl[\Theta(z)-\Theta(z-a)\biggr](\epsilon_{\parallel}-\epsilon_{\perp})$
 such that equation (\ref{epsil0}) reads
 ${\cal E}_0(z)=\epsilon_0(z)\un+\Delta(z)\chat\chat$.

Solutions of equation (\ref{pert-p}) are obtained
 using the Green's functions\cite{hill,celli}.
 There exist, a priori, nine Green's functions and
 thanks to the Dirac distributions appearing in equation (\ref{dirac}), they must only
 be determined for $z'=0$ and for $z'\ge a$ (we remind that we 
are interested by the solution
 in the region $z\gg a$). Furthermore, for $z\gg a$ and $z<0$, all terms 
 of equation (\ref{pert-p}) is front
 of the field components $E'_x({\bf K};z)$ and $E'_y({\bf K};z)$ vanish.
 Hence, the wave equation being expressed as function of the 
 continuous field components, only one non zero Green's function $G_p({\bf K};z,z')$
 exists\cite{hill,celli} in the two relevant regions $z>a'$, $z'\le 0$ and $z>z'$, $z'\ge a$.
 The solution of equation (\ref{pert-p}) therefore reads:
\begin{align}\label{green-int}
 D'_z({\bf K};z)=\int_{-\infty}^{\infty} G_p({\bf K};z,z')
\biggl(
\frac{\partial}{\partial z'}\frac{1}{\epsilon_0(z')}
\biggl[-i{\bf K}\cdot{\bf F}({\bf K};z')+
i\frac{\Delta(z')}{{\cal E}_{zz}(z')}({\bf K}\cdot\chat)c_zF_z({\bf K};z')
\biggr]\nonumber\\
+\frac{K^2}{{\cal E}_{zz}(z')}F_z({\bf K};z')
\biggr) {\rm d}z'\,,
\end{align}
for $z>a$, where the Green's function is given by \cite{maradudin}
\begin{equation}\label{gdef}
G_p({\bf K};z,z')=\frac{1}{W}\biggl(
E_p^<({\bf K};z)E_p^>({\bf K};z')\Theta(z'-z)+
E_p^>({\bf K};z)E_p^<({\bf K};z')\Theta(z-z')
\biggr)
\end{equation}
with
\[
W=E_p^<({\bf K};z)\frac{\partial E_p^>({\bf K};z)}{\partial z}-
E_p^>({\bf K};z)\frac{\partial E_p^<({\bf K};z)}{\partial z}
\]
according to a theorem that can be found in \cite{theorem}. Here
 $E_p^<({\bf K};z)$ and $E_p^>({\bf K};z)$ are the two independent
 plane-wave solutions of the unperturbated equation (\ref{wave-eq0}): $E_p^>({\bf K};z)$
 corresponds to a wave coming from $z\rightarrow -\infty$ and  $E_p^<({\bf K};z)$
 to a wave coming from $z\rightarrow +\infty$. They are thus defined by 
 the following boundary conditions: 
\begin{align}
&\lim_{z\rightarrow +\infty}E_p^>({\bf K};z)\propto \exp(-ik_zz)
\nonumber\\
&\lim_{z\rightarrow -\infty}E_p^<({\bf K};z)\propto \exp(ik_zz)
\,.\nonumber
\end{align}
with $k_z=+(k^2-K^2)^{1/2}$.

Integrating by part the first term in the integral of equation (\ref{green-int}) and
using equations (\ref{Fvec}-\ref{Fvec-fin}), one gets
\begin{align}
 E'_p({\bf K};z)=&\pi\biggl\{
-i{\bf\hat K}\cdot\chat \Delta_n\biggl(1+\frac{1}
{n_o^2(n_o^2+\Delta_nc_z^2)^2}\biggr)
c_z\widetilde{D^p_z}'({\bf K})\nonumber\\
&-i\frac{n_o^4-1}{n_o^2}\biggl({\bf\hat K}\cdot\xhat
\widetilde{E^p_x}'({\bf K};z)+
{\bf\hat K}\cdot\yhat\widetilde{E^p_y}'({\bf K};z) \biggr)\nonumber\\
&-i\frac{\Delta_n{\bf\hat K}\cdot\chat}{n_o^2+\Delta_nc_z^2}
\biggl[1+n_o^2+\Delta_nc_z^2\biggl(1-\frac{1}{n_o^2(n_o^2+\Delta_nc_z^2)}
\biggr)\biggr]\nonumber\\
&\times\biggr(\widetilde{E^p_x}'({\bf K})c_x+
\widetilde{E^p_y}'({\bf K})c_y\biggr)\nonumber\\
&+K\biggl[
1+\frac{1}{(n_o^2+\Delta_nc_z^2)^2}\biggr]
\biggl(\biggl[n_o^2-1+\Delta_nc_z^2\biggr]
\widetilde{D_z}({\bf K};z)\nonumber\\
&+ c_z\Delta_n\biggl[c_x\widetilde{E^p_x}({\bf K};z)+
c_y\widetilde{E^p_y}({\bf K};z)\biggr]
\biggr)
\biggr\}
\label{eprime_p}
\end{align}
for $z>a$ and with $\Delta_n=n_e^2-n_o^2$ and
 ${\bf\hat K}=(K_x\xhat+K_y\yhat)/K$. To obtain this expression,
 we used the definition
 ${\bf E}'_p({\bf K};z)=k/K{\bf E}'_z({\bf K};z)$  
 with ${\bf\hat K}=\yhat$ when $K=0$ \cite{celli}
 (see figure \ref{fig1} for the definition of the reference axes).
 To shorten  equation (\ref{eprime_p}) we also introduced
\begin{align}
\widetilde{D^p_z}({\bf K};z)=&
G_p({\bf K};z,a){\cal F} \biggl[
D_{0z}({\bf R};a)\frac{h_a({\bf R})}{\lambda}\biggr]-\nonumber\\
&G_p({\bf K};z,0){\cal F} \biggl[
D_{0z}({\bf R};0)\frac{h_0({\bf R})}{\lambda}
\biggr]
\label{GE1}\\
\widetilde{D^p_z}'({\bf K};z)=&
\frac{{\rm d}G_p({\bf K};z,z')}{{\rm d}z'}\biggr|_{z'=a}{\cal F}\biggl[
D_{0z}({\bf R};a)\frac{h_a({\bf R})}{\lambda}\biggr]-\nonumber\\
&\frac{{\rm d}G_p({\bf K};z,z')}{{\rm d}z'}\biggr|_{z'=0}
{\cal F}\biggl[D_{0z}({\bf R};0)\frac{h_0({\bf R})}{\lambda}\biggr]
\label{GE2}
\end{align}
where $\lambda$ is the laser wavelength.
Identical expressions hold for $\widetilde{E^p_x}({\bf K})$,
 $\widetilde{E^p_y}({\bf K})$, $\widetilde{E^p_x}'({\bf K})$
 and $\widetilde{E^p_y}'({\bf K})$ . 
  
To derive equation (\ref{eprime_p}) we assumed that \cite{maradudin1}
\begin{equation}\label{prescrip}
\int_{-\infty}^{\infty} f(z)\delta(z){\rm d}z=\frac{1}{2}
\biggl[\lim_{z\rightarrow 0^+}f(z)
+\lim_{z\rightarrow 0^-}f(z)\biggr]
\end{equation}
where $f(z)$ is a discontinuous function, but with a finite jump. Although
 this expression is not mathematically justified as stated in \cite{maradudin1}, it
 can however be used by considering that the Heaviside functions
 of equations (\ref{epsil},\ref{epsil0}) are given by the limit
\[
\Theta(z)=\lim_{\zeta \rightarrow 0} [1+{\rm tanh}(z/\zeta)]/2\,.
\]
This choice is justified by the freedom existing in the
 determination of the dielectric tensor at $z=0$
 \cite{marseilles}. It is to mention that equation (\ref{prescrip}) leads to a disagreement
 with the boundary matching method
 for isotropic-isotropic interfaces in the case of oblique incidence.
Another prescription was proposed in \cite{mills} to avoid this discrepancy. But,
  as mentioned in \cite{maradudin2}, 
 no general proof was provided in \cite{mills}. 
There is then no reason for this particular
 prescription to
 work also for isotropic-anisotropic interfaces. In addition,
 since we are going to restrict ourselves to
 normal incidence, we choose to use the more intuitive and 
 symmetric prescription of equation (\ref{prescrip}) for our calculations.


\subsection{$s$ scattered wave}

Following the lines of the previous section, we get the $s$ wave 
 equation:
\begin{align}
&\biggl(K^2-\omega^2\mu_0\epsilon_0(z)-\frac{\partial^2}{\partial z^2}
\biggr)E'_s({\bf K};z)-\nonumber\\
&\omega^2\mu_0\frac{\Delta(z)}{{\cal E}_{zz}(z)}\shat\cdot\chat
\biggl(\epsilon_0(z)\biggl[c_x E'_x({\bf K};z)+
 c_y E'_y({\bf K};z)\biggr]+c_zD'_z({\bf K};z)\biggr)\nonumber\\
&=\omega^2\mu_0\biggl({\bf F}({\bf K};z)\cdot\shat-
\frac{\Delta(z)}{{\cal E}_{zz}(z)}\chat\cdot\shat c_zF_z({\bf K};z)
\biggr)
\label{eq-swave}
\end{align}
with $E'_s({\bf K};z)=\shat\cdot {\bf E}'({\bf K};z)$ and
 $\shat=(-K_y\xhat+K_x\yhat)/K$. The solution is given by
\[
E'_s({\bf K};z)=\omega^2\mu_0
\int_{-\infty}^{\infty} G_s({\bf K};z,z')\biggl({\bf F}({\bf K};z')\cdot\shat-
\frac{\Delta(z')}{{\cal E}_{zz}(z')}\chat\cdot\shat c_zF_z({\bf K};z')
\biggr){\rm d}z'\,,
\]
with $G_s({\bf K};z,z')$ the $s$ wave Green's function for which an expression
 similar to equation (\ref{gdef}) holds. After integration, one finds 
\begin{align}
E'_s({\bf K};z)=&
\frac{2\pi^2}{\lambda}\biggl\{
(n_o^2-1)\biggl[s_x\widetilde{E}_x({\bf K};z)+
s_y\widetilde{E}_y({\bf K};z)\biggr]
\nonumber\\
&+\frac{\Delta_n}{n_o^2+\Delta_nc_z^2}\shat\cdot\chat\biggl(
n_o^2-\frac{\Delta^2_nc_z^2}{n_o^2+\Delta_nc_z^2}
\biggr)
\biggl[c_x\widetilde{E}_x({\bf K};z)+
c_y\widetilde{E}_y({\bf K};z)\biggr]
\nonumber\\
&+\Delta_n\shat\cdot\chat\biggl(1+\frac{1}{(n_o^2+\Delta_nc_z^2)^2}
\biggr)
c_z\widetilde{D}_z({\bf K};z)
\biggr\}\label{eprime_s}
\end{align}
where $\widetilde{E}^s_x({\bf K};z)$, $\widetilde{E}^s_y({\bf K};z)$
 and $\widetilde{D}^s_z({\bf K};z)$ are obtained by substituting 
 $G_p({\bf K};z,z')$ by  $G_s({\bf K};z,z')$ in equations (\ref{GE1}-\ref{GE2}).

\subsection{Transmitted intensity}

Anticipating the numerical studies of section \ref{numerical}, 
 we shall now consider a Gaussian beam at normal incidence
 coming from the region $z<0$. Expressions for the
 electric field at $z=0$ and $z=a$ read:
\begin{align}
{\bf E}_0({\bf K};0)&={\bf E}_i({\bf K};0)+{\bf E}_r({\bf K};0)\nonumber\\
{\bf E}_0({\bf K};a)&={\bf E}_t({\bf K};a)\nonumber
\end{align}
with\cite{moi}
\begin{align}
{\bf E}_i({\bf K};0)&=\frac{w_0}{\sqrt{2\pi}}\exp\biggl(-\frac{w_0^2K^2}{4}
\biggr)M_{3\times 3}\ezero\label{eq:Et1}
\\
{\bf E}_r({\bf K};0)&=\frac{w_0}{\sqrt{2\pi}}\exp\biggl(-\frac{w_0^2K^2}{4}
\biggr)
\Omega M^>_r\Omega^T M_{3\times 3}\ezero
\\
{\bf E}_t({\bf K};a)&=\frac{w_0}{\sqrt{2\pi}}\exp\biggl(-\frac{w_0^2K^2}{4}
\biggr)
\Omega M^>_t\Omega^T M_{3\times 3}\ezero\label{eq:Et2}
\end{align}
where we chose the beam waist position at $z=0$ and where\cite{moi}:
 $\ezero$ is the electric vector describing the polarisation of the 
 Gaussian beam centre (i.e. ${\bf K}=0$),
 $M_{3\times 3}$ is a $ 3\times 3$ matrix describing the polarisation 
 of the plane waves constituting the Gaussian beam\cite{fourier-vec},
 $M^>_r$ and $M^>_t$ are the 
 Jones matrices describing the reflection and transmission by the
 uniaxial parallel plate (the upper script $>$ indicates that
 these matrices correspond to an incident wave coming from $z<0$).
 The Jones matrices take into account the multiple reflections inside the 
 platelet. They are determined \cite{yeh4x4}
 in the basis $\{\shat,\phat,\khat\}$ and then
 transformed to the basis $\{\xhat,\yhat,\zhat\}$ thanks to the 
 transfer matrix $\Omega$.

The Green's functions are given by
\[
G_m({\bf K};z,z')=
\begin{cases}
(2ik_z)^{-1}\biggl(
\exp(ik_z[z-z'])+M^<_{r_{mm}}\exp(ik_z[z+z'])
\biggr)
;z'\ge a\\
(2ik_z)^{-1}M^<_{t_{mm}}\exp(ik_z[z-z'])
;z'\le 0
\end{cases}
\]
with $m=1,2$ for $s$ and $p$ waves respectively.
 It is to mention that $M^<_r= M^>_r$
 and  $M^<_t= M^>_t$ when the optical axis is in the plane of interface,
 i.e. when $c_z=0$.

From the above expressions, one can compute equations (\ref{GE1}-\ref{GE2})
 and then the $s$ and $p$ scattered fields. In doing so, the following
 kind of Fourier transform is to be evaluated:
\begin{equation}\label{fourint}
{\cal F}\biggl[E_{0x}({\bf R};a)h_a({\bf R})\biggr]=
\int\int E_{0x}({\bf K'};a)h_a({\bf K}-{\bf K'}){\rm d}^2{\bf K'}\,,
\end{equation}
where, because $<h_a>=0$,
 the Fourier transform of the surface profile $h_a({\bf K}-{\bf K'})$
 vanishes when ${\bf K}={\bf K'}$. However, and this is one of the major
 points of this article, since
 $E_{0x}({\bf K};a)\propto\exp(-w_0^2K^2/4)$ at normal incidence,
 then 
 ${\cal F}[E_{0x}({\bf R};a)h_a({\bf R})]$ does not necessarily vanish
 when $K=0$ as it is the case for a
 single plane wave. Consequently, the specular transmitted beam receives a non vanishing
 contribution from the scattered field, even in the first order perturbation
 theory. If the Gaussian beam is viewed as a superposition of plane waves \cite{siegman},
 then this phenomenon is then due to the angular distribution of the plane waves.
 However, this contribution depending on $h_a({\bf K}-{\bf K'})$,
 a realistic simulation of the surface roughness is needed to evaluate the
 integral of equation (\ref{fourint}). This is the subject of the next section.

To exhibit this contribution, let us assume that
 a Wollaston prism is located after the anisotropic plate and that its 
 axes correspond to the $\xhat$ and $\yhat$ directions.
 Writing the scattered electric field in the basis $\{\xhat,\yhat,\zhat\}$,
\begin{align}
E'_x({\bf K};z)&=
\Omega_{11} E_s'({\bf K};z)+\Omega_{12}E_p'({\bf K};z)\nonumber\\
E'_y({\bf K};z)&=
\Omega_{21} E_s'({\bf K};z)+\Omega_{22}E_p'({\bf K};z)\nonumber
\end{align}
one obtains
\begin{align}
I_x&=\int\int|\xhat\cdot{\bf E}_t({\bf K};z)|^2{\rm d}^2{\bf K}+
\int\int|E'_x({\bf K};z)|^2{\rm d}^2{\bf K}+
\nonumber\\
&\int\int \biggl[\xhat\cdot{\bf E}_t({\bf K};z)\biggl(E'_x({\bf K};z)\biggr)^*+
 \biggl(\xhat\cdot{\bf E}_t({\bf K};z)\biggr)^*E'_x({\bf K};z)\biggr]
 {\rm d}^2{\bf K}
\nonumber\\
I_y&=\int\int|\yhat\cdot{\bf E}_t({\bf K};z)|^2{\rm d}^2{\bf K}+
\int\int|E'_y({\bf K};z)|^2{\rm d}^2{\bf K}+\nonumber\\
&\int\int \biggl[\yhat\cdot{\bf E}_t({\bf K};z)\biggl(E'_y({\bf K};z)\biggr)^*+
 \biggl(\yhat\cdot{\bf E}_t({\bf K};z)\biggr)^*E'_y({\bf K};z)\biggr]
 {\rm d}^2{\bf K}
\nonumber
\end{align}
for the two intensities $I_x$ and $I_y$ measured after the Wollaston.
 Writing $I_x=I_x^{[0]}+I_x^{[1]}+I_x^{[2]}$, and $I_y$ in the same way, with
\begin{align}
I_x^{[0]}=&\int\int|\xhat\cdot{\bf E}_t({\bf K};z)|^2{\rm d}^2{\bf K} \label{IX}\\
I_x^{[1]}=&\int\int \biggl[\xhat\cdot{\bf E}_t({\bf K};z)\biggl(E'_x({\bf K};z)\biggr)^*+
 \biggl(\xhat\cdot{\bf E}_t({\bf K};z)\biggr)^*E'_x({\bf K};z)\biggr]
 {\rm d}^2{\bf K} \\
I_x^{[2]}=&\int\int|E'_x({\bf K};z)|^2{\rm d}^2{\bf K} \label{IY}
\end{align}
one sees that the specular-scattered interference term $I_x^{[1]}$
 is of first order in $\sqrt{<h^2>}/\lambda$. 
 As $w_0\rightarrow\infty$, the Fourier transforms of the electric
 field components lead to Dirac distributions and the usual plane wave 
 result is recovered. This interference term is therefore expected to
 be dependent on the laser beam waist.

 Let us finally note that equation (\ref{IY}) does not completely describe the second 
 order contribution $I_x^{[2]}$ in the specular region, the interference between the 
 specular and the second order scattered field being omitted in our calculation.

\section{Numerical simulations}\label{numerical} 
To estimate the specular-scattered
 interference term, a laser beam crossing a
 quartz platelet at normal incidence is now considered. The incident
 electric vector is fixed to $\ezero=\xhat$, i.e. perpendicular to the 
 plane of incidence of the Gaussian beam's centre, and the intensities recorded
 after a Wollaston prism are calculated as in the previous section.

\subsection{Numerical ingredients and input parameters}

As described in the previous section, the specular-scattered
 interference term can only be evaluated if a simulation of the surface roughness is provided.
 The profiles of
 the quartz surfaces $h_0$ and $h_a$ are thus randomly generated independently.
 Then the scattered fields are computed
 for various orientations of the optical axis and the two intensities
  of equations (\ref{IX}-\ref{IY}) are determined. The procedure is repeated 
 in order to obtain a statistical distribution of the intensities. 
The numerical integrations of 
 equations (\ref{IX}-\ref{IY}) are performed
 in the domain $\arctan(K/k)<1^o$, which roughly
 corresponds to the angular acceptance of a 1~inch diameter optical lens  
 located at 1~m from the plate. In practice this limit does not affect 
 the value of the interference term but only the second order contribution.

 Random profile generations are performed as follows.
 First, the height 
 distributions are determined according to a Gaussian distribution
 of mean value zero and root mean square $\sigma=6\,{\rm\AA}$. This is 
 a typical value for a high quality polished quartz plate \cite{quartz}. 
The heights are stored in a grid $\{x,y\}$ of length $L\times L$
 containing $2^{n}\times 2^{n}$ nodes. The value of $n$ is limited
 by the computer memory capacity, $n=11$ in our case. Next, the Fourier transform is 
 computed and the profile heights are weighted by the square root of the 
 two dimensional spectral density function\cite{bennett} (PSD2). Though we are going to
 consider an isotropic roughness distribution, we shall not use the reduced one dimensional 
 radial PSD1\cite{bennett,taback}. In doing so we fully account for the random nature
 of the surface roughness in the evaluation of the specular-scattered interference term (the use
 of a PSD1 would leads to a smaller dispersion of our numerical results).

 The result of \cite{quartz} for the PSD2 is used:
\begin{equation}\label{eq:ps2d}
PSD2(K)=
\begin{cases}
2\pi\sigma^2l^2 (1+K^2l^2)^{-3/2}\,;\, K\ge K_{min}\\
0\,;\, K<K_{min}\\
\end{cases}
\end{equation}
 where the cut-off spatial frequency $K_{min}$ has been introduced to account for the
 surface profile property $<h>=0\Rightarrow PSD2(0)=0$ \cite{fractal}. The 
 correlation length $l$ is of the order of\cite{quartz} $1\mu$m  and
 $K_{min}$ is smaller than $10^{-3}\mu$m$^{-1}$ \cite{surf}. The
  parameter $1/K_{min}$ acts as a spatial frequency threshold for the laser
 radius: roughly speaking,
 for $w_0>\sqrt{2}/K_{min}$ the Gaussian beam behaves as a
 plane-wave and the specular-scattered interference term vanishes.
 Nevertheless, the present values of $K_{min}$
 are limited by the acceptance of the surface profile
 measurements:
 values as small as $10^{-4}\mu$m$^{-1}$ for Si substrates\cite{fractal} and
 $10^{-5}\mu$m$^{-1}$ for $K_{min}$ 
 for Si wafers \cite{marx} have been reported (notice that these numbers
 lead to a laser radius threshold greater than 15~mm). Finally, the inverse
 Fourier transform is computed leading to a `coloured' random surface.

 The grid parameters are determined by the correlation length $l$ and
 the laser waist $w_0$. The distance between two nods of the grid
$\{x,y\}$ must be at least twice smaller than $l$ and the distance
 between two nods of the grid $\{K_x,K_y\}$ ten times smaller than
 $\sqrt{2}/w_0$ approximately. This leads to the following  
`experimental' parameters:
 $w_0=100\mu$m, $l=1.6\mu$m and $L=8w_0$.
 As for the laser wavelength we choose $\lambda=0.6328\,{\rm \mu m}$ 
 and $n_0=1.542637$ and $n_e=1.551646$.
 The ratio $\sigma/\lambda$ is therefore
 of the order of $10^{-3}$ in our numerical examples. 

 Using the numbers given above, we have written a computer program to estimate the
  specular-scattered
 interference term. Calculations of the unperturbated electric
 fields and of the Green's functions are performed according to \cite{moi}.
 A fast Fourier transform (FFT) algorithm is used for the simulation of the surface roughness
 and for the integrals of the kind of equation (\ref{fourint}).
 The numerical precision for the unperturbated intensities
 is of the order of $10^{-6}$\cite{moi}. The specular-scattered
 interference term is thus known to an accuracy better than
 $\sigma/\lambda\times 10^{-6}\approx 10^{-9}$ (with our grid size, the FFTs do not reduce
 this accuracy). However, the CPU-time is quite sizable: with a
 SPECfp2000 1288 computer, the  
 random generation of the surfaces takes 22~s and next the computation of the intensities for
 one given orientation of the optical axis takes 183~s.

\subsection{Numerical results}

\begin{figure}[h]
\begin{center}
\epsfig{file=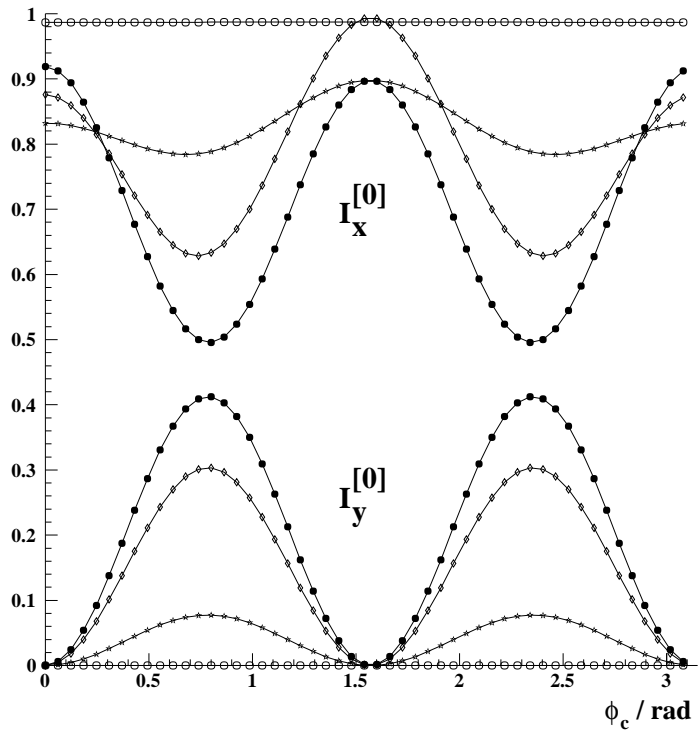,width=10cm}
\end{center}
\vspace{-5mm}
\caption{ Intensity of the specular beams as function of the optical axis azimuth for various
 quartz plates: a tenth order quarter wave plate (full dots), 0.562~mm thickness (open
 dots), 5~mm thickness (diamonds) and a tenth order plate thickness but with the optical
 axis polar angle fixed to $\theta_c=\pi/4$ (stars).
 The upper set of curves represents the intensities $I^{[0]}_x$
 and the lower set of curves $I^{[0]}_y$.   
    \label{fig-I0}}
\end{figure}

 We first consider a quartz plate thickness $a=0.720\,$mm with the optical
 axis located in the plane of interface ($\theta_c=\arccos(c_z)=\pi/2$),
 that is a tenth order
 quarter-wave plate. $I_x^{[0]}$ and $I_y^{[0]}$  are shown is figure \ref{fig-I0} as function of 
 the optical axis azimuth $\phi_c=\arctan(c_y/c_x)$.

\vspace{-5mm}
\begin{figure}[h]
\begin{center}
\epsfig{file=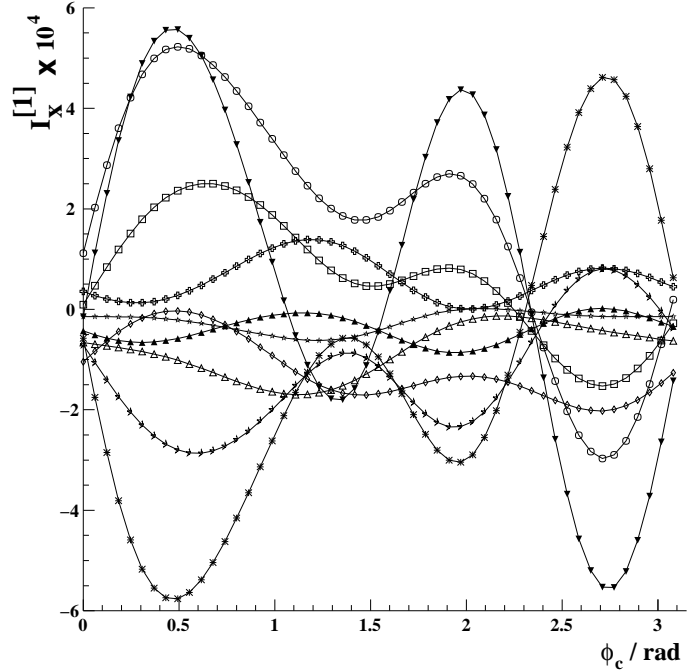,width=9.8cm}
\end{center}
\vspace{-0.5cm}
\caption{ Interference between the specular and scattered transmitted fields $I^{[1]}_x$ as function of 
 the optical axis azimuth. The plate is a tenth order quarter wave plate and the beam 
 waist is $w_0=100\,\mu$m. Different symbols correspond to different random generated
 surface profiles.
    \label{fig-I1x}}
\end{figure}

\begin{figure}[h]
\begin{center}
\vspace{-0.5cm}
\epsfig{file=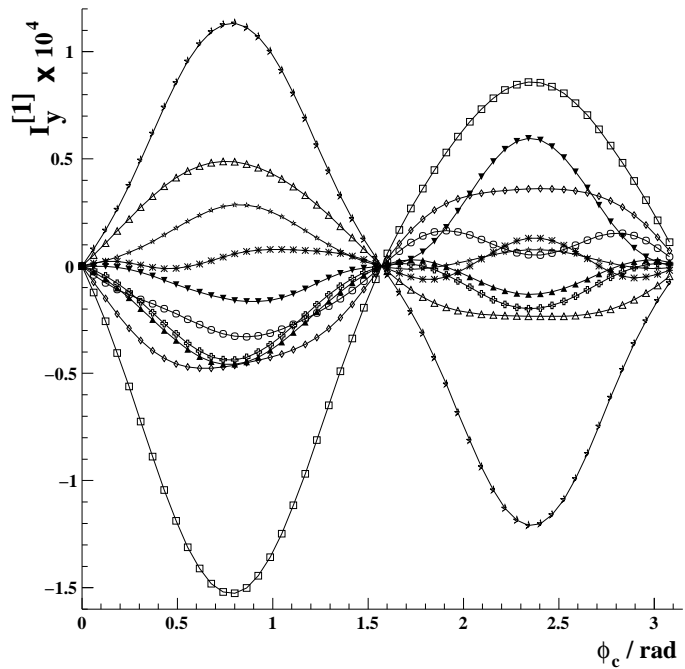,width=9.8cm}
\end{center}
\vspace{-0.5cm}
\caption{  Interference between the specular and scattered transmitted fields $I^{[1]}_y$ as function of 
 the optical axis azimuth. The plate is a tenth order quarter wave plate and the beam 
 waist is $w_0=100\,\mu$m. Different symbols correspond to different random generated
 surface profiles.
    \label{fig-I1y}}
\end{figure}

\newpage
The results for the first order contributions $I_x^{[1]}$ and $I_y^{[1]}$ are shown in
  figures \ref{fig-I1x},\ref{fig-I1y}.\\

  Each curve of these plots corresponds to different
 surface profiles. Considering one given profile, one can notice that: the size of
 the specular-scattered
 interference strongly depends on the surface profile and
  can reach the per mill level of the zero order contribution, its sign
 changes with $\phi_c$ and its shape is not regular with $\phi_c$. The change of sign is expected
 since the intensity averaged over a large number of profiles obviously vanishes. The
 erratic shape is also expected since the fields change with $\phi_c$ and so do 
 the Fourier transforms as the one of equation (\ref{fourint}). 

 The second order 
 contribution (calculated from equation (\ref{IY})) is six order of magnitude smaller
 that the zero order contribution. However, we do not show any results
 since our second order calculation is not complete concerning the specular angular range. 

\begin{figure}[htbp]
\begin{center}
\epsfig{file=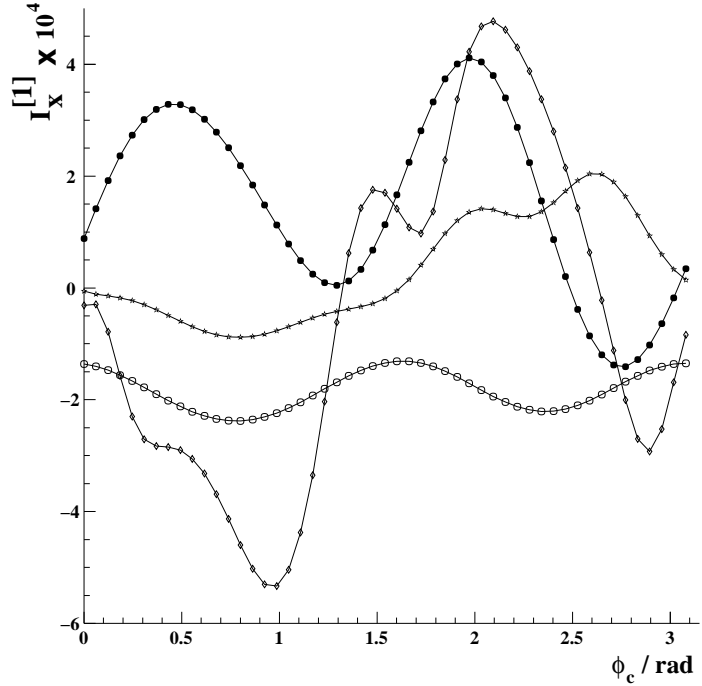,width=10cm}
\end{center}
\caption{ Interference between the specular and scattered transmitted fields $I^{[1]}_x$ as function of 
 the optical axis azimuth for: a tenth order quarter wave plate (full dots),
 0.562~mm thickness (open dots),
  5~mm thickness plate (diamonds) and a tenth order plate thickness but with the optical
 axis polar angle fixed to $\theta_c=\pi/4$ (stars). The surface profiles are the same for the
 four curves.
    \label{fig-eI1x}}
\end{figure}

\begin{figure}[h]
\begin{center}
\epsfig{file=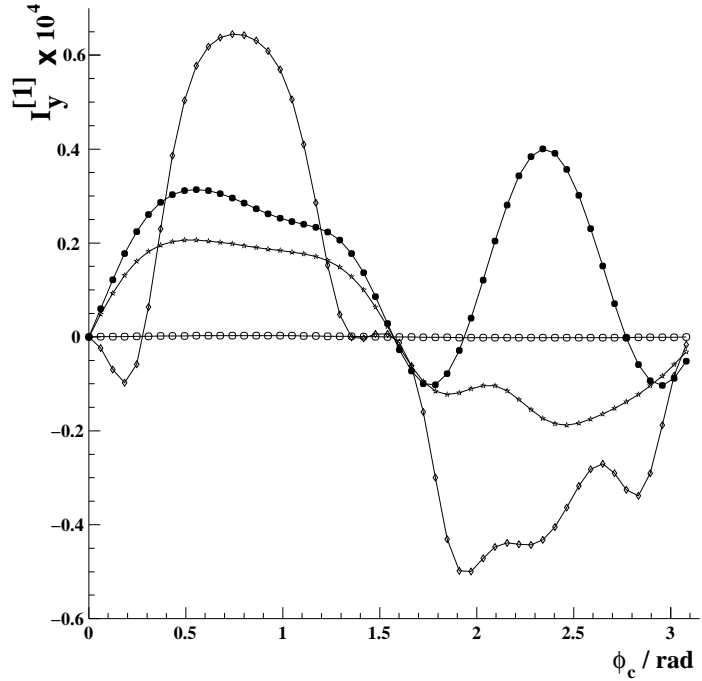,width=10cm}
\end{center}
\caption{Same as figure \ref{fig-eI1x} but for $I^{[1]}_y$.  
    \label{fig-eI1y}}
\end{figure}

 Large differences are indeed
 observed when the plate thickness is changed. The specular-scattered
 interference contributions are computed for 
 $a=0.562$~mm as in \cite{folcia} (i.e. $(8+10^{-3})\times 2\pi$ retardation
 plate with our choice for the optical indices)
and $a=5$~mm as in \cite{simon} (i.e. $(71+0.18)\times 2\pi$ retardation plate),
 and still with $\theta_c=\pi/2$. 
They are compared to the values obtained with the 
 tenth order quarter wave plate and the same surface profiles. 
\newpage
\noindent
 The results are presented in  figures \ref{fig-eI1x},\ref{fig-eI1y}. $I_x^{[1]}$ and
 $I_y^{[1]}$ scale 
 with $I_x^{[0]}$ and $I_y^{[0]}$ (see figure \ref{fig-I0}). In particular, the oscillations
 of $I_x^{[1]}$ are dumped when  $I_x^{[0]}$ gets flat (i.e. for the almost zero 
 retardation plate $a=0.562$~mm).

To investigate the dependence of equations (\ref{eprime_s},\ref{eprime_p})
 on the optical axis polar angle
 $\theta_c$, the calculations were performed fixing $\theta_c=\pi/4$ for
 the plate thickness $a=0.720\,$mm. Here again the 
 variations are noticeable (see figures \ref{fig-eI1x} and \ref{fig-eI1y}). 

 Looking at  figures \ref{fig-I1y} and \ref{fig-eI1y}, one can remark that  
   $I_y^{[1]}$ tends to be of opposite sign in the regions $0<\phi_c<\pi/2$
 and $\pi/2<\phi_c<\pi$. But this is not a general rule as it seems to 
come out from experiments \cite{folcia,simon}. One can also observe two fix points at
  $\phi_c=0$ and $\pi/2$ on figures \ref{fig-I1y} and \ref{fig-eI1y}. $I_y^{[1]}$ being the interference
 between the scattered field and the zero order field, these fix points correspond 
  to the zeros of the zero order field (see  $I_y^{[0]}$ on figure \ref{fig-I0}). 
 This is not the case for the second order
 term of equation (\ref{IY}) which is of the order of $10^{-6}$ and therefore dominates around 
 $\phi_c=0,\pi/2$ (here the missing term of equation (\ref{IY}) is not relevant
 since it describes the interference between the specular and the second order scattered fields).
 However, the dispersion of $I_y^{[1]}$  
around zero for $\phi_c=0,\pi/2$ (not visible on these figures)
 defers very slightly from zero, it is of the order
 of $10^{-10}$ for $w_0=100\,\mu$m and $\approx 10^{-8}$ for $w_0=25\,\mu$m.
 This is a cross-polarisation effect,
 i.e. this is due to the matrix $M_{3\times 3}$ in equations (\ref{eq:Et1}-\ref{eq:Et2}).

\newpage

 The numerical results presented here are rather independent of the 
 choice for the PSD2 provided a quartz plate of optical grade is considered. It is indeed
 experimentally demonstrated \cite{surf,bennett,taback,fractal,marx}
 that the PSDs of optical element's surface have an inverse-power-low (or Fractal-like) behaviour.
 Therefore, various smooth mathematical
 representations of the PSD (see \cite{elson,rasigni} for examples) are reducible 
 to equation (\ref{eq:ps2d}) as it is justified in \cite{taback}.

As a concluding remark, we point out that three important dimensionless parameters
 $\omega_0/l$, $\sigma/\lambda$
 and $l/\lambda$
 have been encountered in our calculations ($\omega_0/\lambda$ describes
 the cross-polarisation effects discussed above and is therefore
 not relevant here).

 As mentioned in section \ref{theory},
 the validity of the perturbation treatment depends on 
  $\sigma/\lambda$ and $l/\lambda$. Using the first order perturbation theory,
 the severe conditions $\sigma/\lambda\ll 1$ and $l/\lambda\simeq 1$
 must hold\cite{acoustic}. They are fortunately fulfilled by optical grade quartz plates. 
 To study the influence of the correlation length $l$, we changed the value of $l$
 to $0.7\mu$m and $2\mu$m and we observed no significant qualitative
 differences with respect to the results described above.
\newpage
 As for the last dimensionless parameter
$\omega_0/l$, we already mentioned that when $\omega_0/l\rightarrow\infty$ the usual result for
 plane waves is recovered (i.e. the  specular-scattered
 interference term vanishes) although, with regard to the
 cut-off parameter $K_{min}$,
 this limit seems to be idealistic for a finite size quartz plate. To get an idea of the 
 influence of $w_0$, we increased it to $200\mu$m and here again, no significant differences
 were observed. Much larger values for $w_0$ could not be tried, keeping a reasonable
 correlation length, because of the computer memory limitation. Finally let us mention that 
 the other limit $\omega_0\ll l$ corresponds to the scattering by gratings \cite{francais}.
 In this limit the
  specular-scattered interference term vanishes since the diffusion occurs at large angle
 with respect to the specular beam direction.

\section{Conclusion}

We have computed, in the leading order perturbation theory,
 the effect of surfaces roughness on uniaxial platelets transmittance. Taking
 into account the Gaussian nature of laser beams we showed that the interference
 between the specular and scattered fields contributes
 to the intensity measurement performed in the specular region.

 This contribution is of first order
 in the ratio of the root mean square roughness over the laser wavelength $\sigma/\lambda$.
 It depends strongly
 on the plate surfaces profiles and on the
 crystal optical properties, orientation of the optical axis, thickness
 and optical indices (i.e. temperature). It is therefore useless to implement the 
 roughness calculation in a HAUP type of fitting procedure (in addition, the numerical
 calculation are computer time consuming).

 In view of our
 numerical results, it is most likely
 that simple overlayer models \cite{folcia} cannot describe
 accurately
 the dynamical properties of our main formula equations (\ref{eprime_p},\ref{eprime_s}). 
 Nevertheless, we point out that, because of the random property of the specular-scattered
 interference term, a simple way to avoid it is to perform a series of measurements at different
 locations on the plate and then to average the results. Although this procedure
 would increase the 
 uncertainty on the determination of crystal optical
 parameters, it should however decrease the systematic bias. The determination of the 
 plate thickness in situ could be done by varying the laser incident angle (i.e. by tilting the
 plate) \cite{bretagne}.

\subsection*{Acknowledgement}
I would like to thank J.P. Maillet for suggestions and enlightening discussions.
 I would also like to thank M.A. Bizouard for helpful discussions and F. Marechal for careful reading.

\end{document}